\documentclass[12pt]{article}

\advance\voffset by -0.25cm

\usepackage{graphicx,amsmath,amsfonts,amsthm,bbm,setspace,url}
\usepackage{natbib,verbatim,mathrsfs,float,epstopdf,moreverb,multicol}
\usepackage{lscape}

\DeclareGraphicsExtensions{.pdf,.png,.jpg}

\newcommand{\be}{\begin{enumerate}}
\newcommand{\ee}{\end{enumerate}}

\begin{document}
	\sloppy
	\baselineskip=20pt
\begin{center}{
		\Large \bf
		Mapping the uncertainty of 19th century West African slave origins using a Markov decision process model.
	}
	
	\bigskip
	\bigskip
	
	{\bf
		Zachary Mullen\footnote[1]{Department of Applied Mathematics,
			Author e-mail:
			\texttt{zachary.mullen@colorado.edu}}
	  and 
		Ashton Wiens\footnote[2]{Department of Applied Mathematics,
				University of Colorado, Boulder, CO.}
	and
		Eric Vance\footnote[3]{Department of Applied Mathematics,
			University of Colorado, Boulder, CO.}
	and
		Henry Lovejoy\footnote[4]{Department of History,
			University of Colorado, Boulder, CO.}
	}
	
	\bigskip
	\bigskip
	
	{\bf \today}
	
\end{center}

\bigskip
\bigskip

\begin{abstract}
The advent of modern computers has added an increased emphasis on channeling computational power and statistical methods into digital humanities.  Including increased statistical rigor in history poses unique challenges due to the inherent uncertainties of word-of-mouth and poorly recorded data.  African genealogies form an important such example, both in terms of individual ancestries and broader historical context in the absence of written records.  Our project aims to bridge the lack of accurate maps of Africa during the trans-Atlantic slave trade with the personalized question of where \textit{within} Africa an individual slave may have hailed.  We approach this question with a two part mathematical model informed by two primary sets of data.  We begin with a conflict intensity surface which can generate capture locations of theoretical slaves, and accompany this with a Markov decision process which models the transport of these slaves through existing cities to the coastal areas.  Ultimately, we can use this two-step approach of providing capture locations to a historical trade network in a simulative fashion to generate and visualize the conditional probability of a slave coming from a certain spatial region given they were sold at a certain port. This is a data-driven visual answer to the research question of where the slaves departing these ports originated.
\end{abstract}

\section{Introduction}
The study of colonial empires is dominated by incomplete data and missing records.  Despite this difficulty, there is significant interest in tracing the forced diaspora of African peoples via slavery \citep{lovejoywebsite}.  Many black organizations in the modern Americas can trace their origins to the cultural unity required in overcoming the struggles of their subjugation \citep{chambers2012}.

To date, much of the work in understanding this cultural genesis has focused on genealogy and literary interpretation, but the forced relocation of a predominately illiterate population leads to significant shortcomings in availability of written history.  As a result, modern understanding of the exodus often lacks comprehensive regional descriptions of the socio-political climate within Africa that enabled the internal slave trade then exploited by colonial European powers.  The growing field of digital humanities attempts to expand upon logocentric analyses of African history with modern methods in text mining, linguistic analysis, and machine learning \citep{lovejoytalk2017}. These GIS and geospatial methods \citep{knowles2008placing} have been employed heavily in World War II and Holocaust studies \citep{knowles2014geographies}.

One West African slave-trading state was the Oyo empire, which peaked in the late 18th century, culminating in a rapid decline over a series of crises and invasions around the 1820s.  During these conflicts, slavers regularly departed from the coast of the Oyo empire and bordering West African states, and many of these voyages are well documented by the slave traders.  In addition to ship logs, a handful anecdotal accounts of individual slave movements from the collapse of the Oyo empire have been reconstructed from written and oral records \citep{kelley2016origins,kelley2016voyage} (maybe cite slavebiographies.org).  Recent work has emphasized integrating the collapse of the Oyo empire into the digital humanities, including the creation of detailed maps on the shifting borders of the collapsing empire \citep{lovejoy2013redrawing}.

One current question regarding the collapse of the Oyo empire is exploring the logistics and detailed movements of the internal slave trade and how those systems actually filled the ships leaving the West African coast.  In many cases, the state-controlled ships have accurate passenger counts, ports of arrival, and ports of origin, and modern genealogical explorations can often trace ancestries to those specific ships. However, little historical evidence explicitly connects the passenger logs - where available - and the movements of the ships to the politics of inland Africa at the time.  Questions of ancestry often dead-end at these transit points despite the work and literature documenting the internal conflicts during the Oyo collapse.  

We attempt to expand on the understanding of the internal slave trade of the Oyo Empire by synthesizing spatial mathematical models onto conflict maps and conjoining them with models for decision processes governing inland slave movements.  The first question is one of using discrete events such as recorded dates of battles or towns destroyed to create a model for the location and intensities of conflict.  We use spatial smoothing on  recorded conflict events to create a continuous density map of the warring regions, augmenting the existing maps of shifting borders by an accompanying picture of which cities and regions in the empire were most likely locations for slavers to capture individuals.  We couple this map of conflict regions with a Markov decision process for the Oyo region's internal slave trading network.  We view adjacent or nearby cities as a connected network, and the Markov decision process attempts to ask: "what are the likely movement paths" of slaves captured until their eventual sales and departures via ship or into the trans-Sahara region.

The goal is to provide a functional and descriptive model for the most likely inland origin locations of slaves given a known year and port of origin. As a result, the conflict map and slaver decision process models combine to answer this: we use the conflict map to generate annual maps of likely locations slaves were captured, then pass them into the trading network to determine where slaves captured at those locations would be most likely to leave the region.  The resulting counts allow for the inverse question as well; e.g. "for all slaves leaving Lagos in 1824, from which conflict regions did they originate?"  This allows our analysis to bridge the process-focused models that stay true to historical narrative with the ends-oriented goals of a genealogist, who may wish to reverse-engineer the historical origin stories.  We hope for our exploration to be applicable and available to historians in both other regions of the African diaspora and to studies of other instances of forced transit, such as the Holocaust or the relocation of American indigenous peoples.

\section{Data}
\label{S:Data}
\textbf{Section in Progress pending collaboration: Describe the conflict data - what historical accounts were used?}

We have several geopolitical data sets, describing the trade routes and conflicts that were we think were present during the collapse of the kingdom of Oyo from approximately 1816-1836 near modern day Togo, Benin, and western Nigeria. The data are shown in \ref{f:1}. For each year, we also have approximations of the total number of slaves departing the region as a whole and specific trading ports. The data were collected in \ref{Mapping the Collapse of Oyo}. 

The conflict data is a table where each row describes a 2D spatial location where a conflict occurred. There are variables describing the start year and end year, as well as the intensity of the conflict. The intensity was encoded as a categorical variable with four levels: 0 means a city is founded, 1 means a city is rebuilt, 5 means a city is attacked, and 10 means a city is destroyed. We did not use the founded/rebuilt city data.

\begin{figure}[t!]
	\centering
	\includegraphics[width=0.95\linewidth]{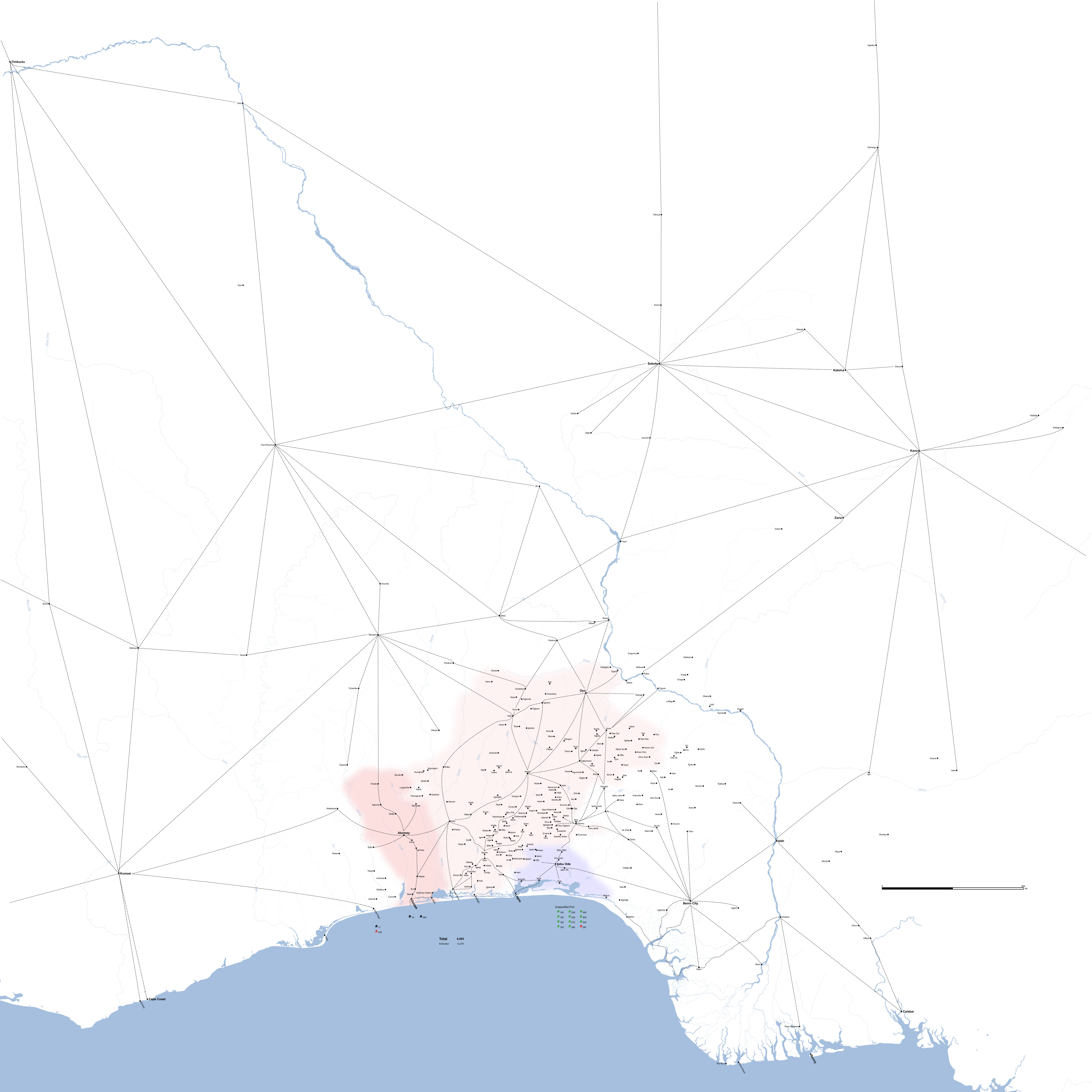}
	\caption{Map of Trade in Oyo, 1816}
	\label{fig:1816TradeMap}
\end{figure}

Similarly, we have a list of cities with spatial coordinates and the years the city existed (dependent on being destroyed or rebuilt). To infer the trade network among these places, we relied on the map \ref{fig:1816TradeMap}.  This map has been informed by both available historical records from the time (\textbf{CITATION MISSING}) and geographic ease of transit between cities.  This representation adds a layer of detail and geopolitical information on top of those published in prior works, such as \citep{lovejoy2013redrawing}.

We encoded the relationships (edges) between the nodes of this graph into an adjacency matrix, describing which cities are connected. An adjacency matrix $A$ for a set of locations (nodes) $s_1, \cdots, s_n$ is of dimension $n \times n$. An entry $A_{ij}$ is nonzero (usually 1) if there is a connection starting at $s_i$ and ending at $s_j$. This formulation describes a directed graphical structure. If the edges are undirected, then $A_{ij} = A_{ji}$ and so $A$ is symmetric. We use this adjacency matrix to construct the probability transition matrix needed in the Markov Decision Process, described in Section \ref{SS:MDP}.

The third data set we have is the port total data: for each year, the total number of slaves leaving each port was estimated using digitally transcribed hand-written ship logs. Some of the estimates are assigned to an unknown port. This data was not used in formulation of any models we develop in this paper, but we used it as validation data to tune parameters in the model.

Finally, we have shapefile data with prominent geographical features that existed in the region during the historical period. In particular, we include bodies of water in plots which are relevant to identifying the boundaries of the various states. \textbf{The data was downloaded from ....}. Several bodies of water which were created since the historical period were removed from the data set.

\section{Model}

\subsection{Mapping Conflicts}\label{SS:krig}

The historical narrative surrounding the fall of the Oyo empire is one of borders collapsing inwards from the independence of Dahomey and lost conflicts to Ilorin and Ijebu.  While the borders of the resulting countries are relatively well known, the question of slave origins requires unpacking the conflicts themselves to determine the regions within the greater Oyo area most impacted by each conflict.  The data available provides conflicts and a measure for their intensity, with battles marked as less intense than the complete destruction of towns.  However, a conflict between nations does not unfold over just the sites of the major battles and events: armies mobilize, raid, and occupy villages throughout contested regions.  To account for this behavior not having explicit historical accounts, we decided to create a continuous spatial map of the conflict borders given the discrete time and place data available.

\subsubsection{Gaussian Process for Origin Locations}
We model our conflicts as arising from a spatially correlated Gaussian Process with an underlying M\a'tern covariance.  Notationally, if we have conflict intensity measures $\boldsymbol{Y}$ observed at 2-D spatial locations $\boldsymbol{S}=\{s_1, s_2, \dots s_n\}$, the Gaussian process considers $\boldsymbol{Y}$ to be a single draw from a multivariate normal on $\mathbb{R}^n$.  This corresponds to a log-density of

$$f(\boldsymbol{Y}) \propto -\log\det\left(\Sigma+\tau^2I\right)-\boldsymbol{Y}^T \left(\Sigma+\tau^2I\right)^{-1} \boldsymbol{Y} $$
 where $\Sigma_{i,j}$ is given by the M\'atern covariance: $
  \Sigma_{i,j}=\sigma^2 k(s_i,s_j)+\mathbbm{1}_{\{i=j\}}(\tau^2)$ for $k(s_i,s_j)=\frac{2^{1-\nu}}{\Gamma(\nu)} 
  \left(a\|s_1-s_2\|\right)^\nu K_\nu(a\|s_1-s_2\|)$
with $K_\nu$ a modified Bessel function of the second kind of order $\nu$ and $\|\cdot\|$ denoting Euclidean distance.

The formal kriging estimator fills in a map of the Oyo region at chosen resolution by taking each desired locations $s_0$ on the fine grid and computing $$\hat{\boldsymbol{Y}}(s_0)=\sigma^2 k(s_0,\vec{s}) \left(\Sigma+\tau^2I\right)^{-1} \boldsymbol{Y}$$

\subsubsection{Alternatives Considered}
There are a variety of alternative mathematical options available to bridge a set of discrete spatial observations (sites of conflict) into a smoother continuous map.  These include smoothing over observed conflicts onto other locations via splines or the above kriging estimators, treating conflicts as draws from an underlying density and using a kernel density estimator (KDE) to recover that density, or viewing conflict sites as actualizations of an inhomogeneous Poisson point process and estimating the associated intensity function.  Of these options, we chose to use a classical Kriging estimator of an underlying Gaussian process for a few reasons.  These include:

\begin{itemize}
	\item A typical Kriging formulation views the data as part of a demeaned autocorrelated process, where the assumption of zero mean rapidly pushes the surface to zero-valued when far from observations \citep{cressie1992statistics}.  This corresponds to the idea that the conflicts and attacked towns themselves were the predominant sources of slaves at the time.
	\item The parameters available to Kriging covariance models are both flexible and can be interpreted in the units of the data.  Specifically, the variance parameter (or sill) is a scaling of the relative importance of minor/major conflicts and the range parameter measures the distances from observed conflicts at which the overall region of conflict exists.
	\item Where a classical kernel density estimate is  symmetric due to its equal weighting of observations-as-kernels, the choice of surface smoothness parameter in a correlated Gaussian Process allows for ridge-like structures that approximate the shifting borders of a conflict region.
\end{itemize}

\begin{figure}[t!]
	\centering
	\includegraphics[width=0.3\linewidth]{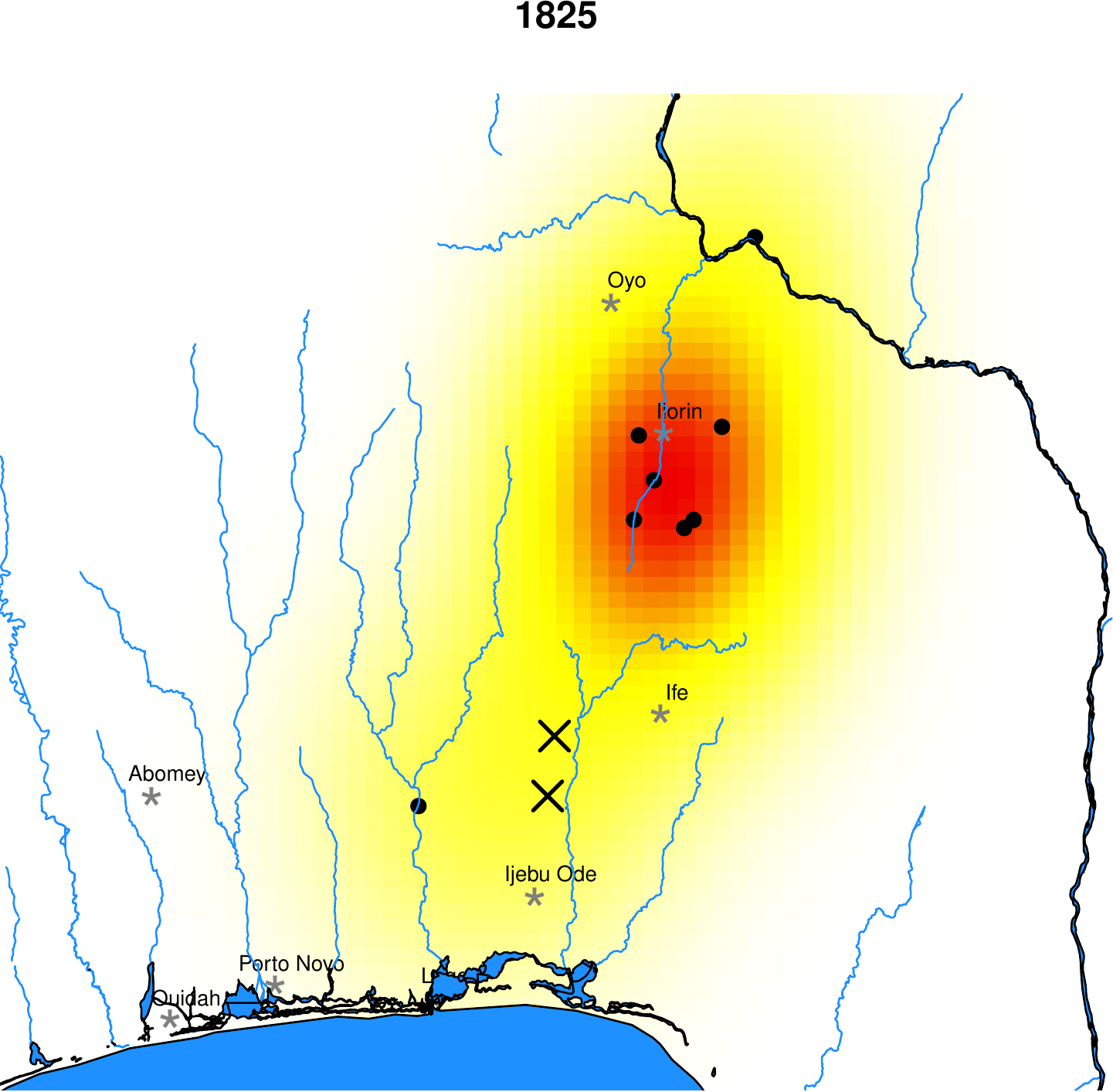}
	\includegraphics[width=0.3\linewidth]{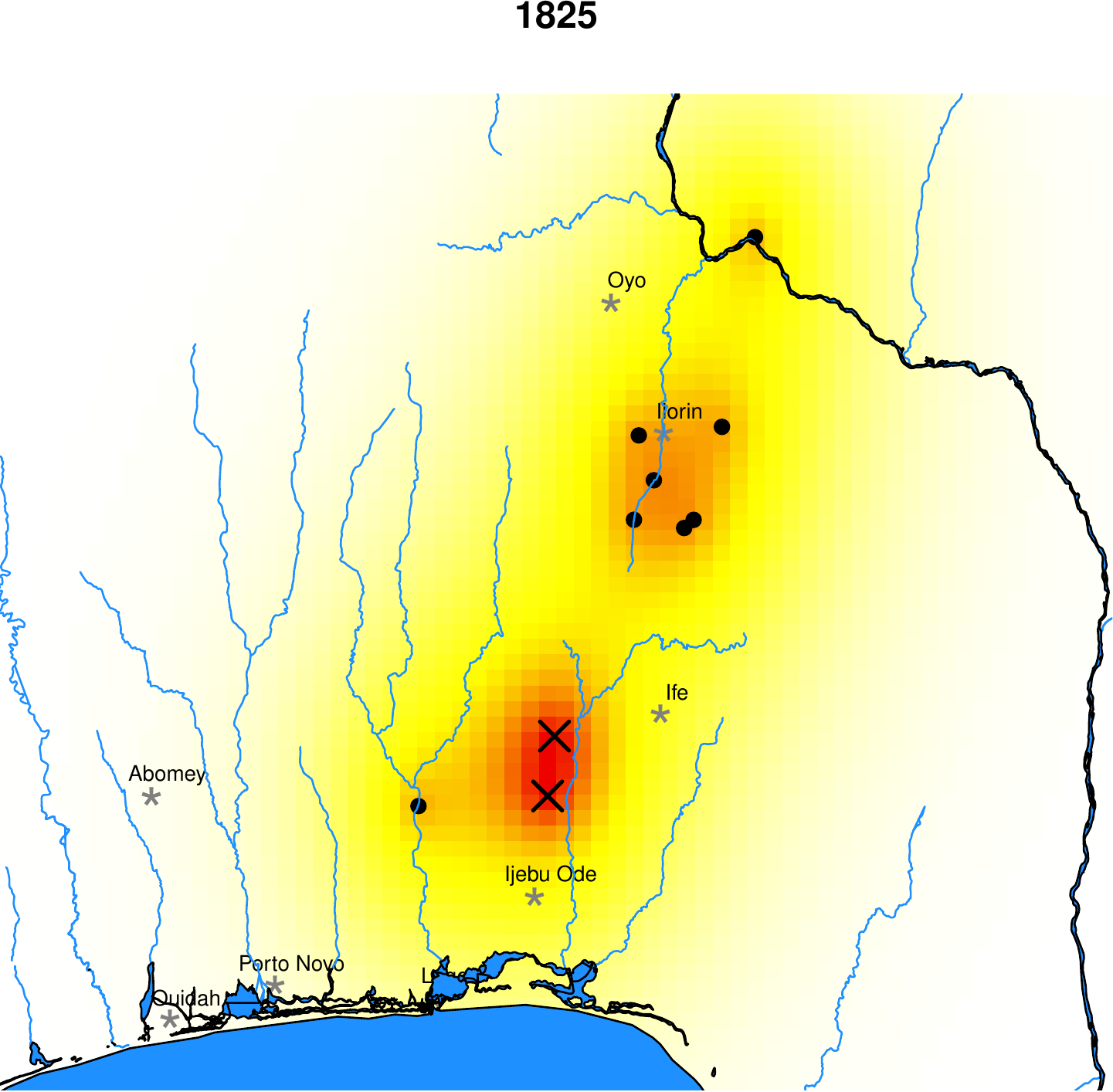}
    \includegraphics[width=0.3\linewidth]{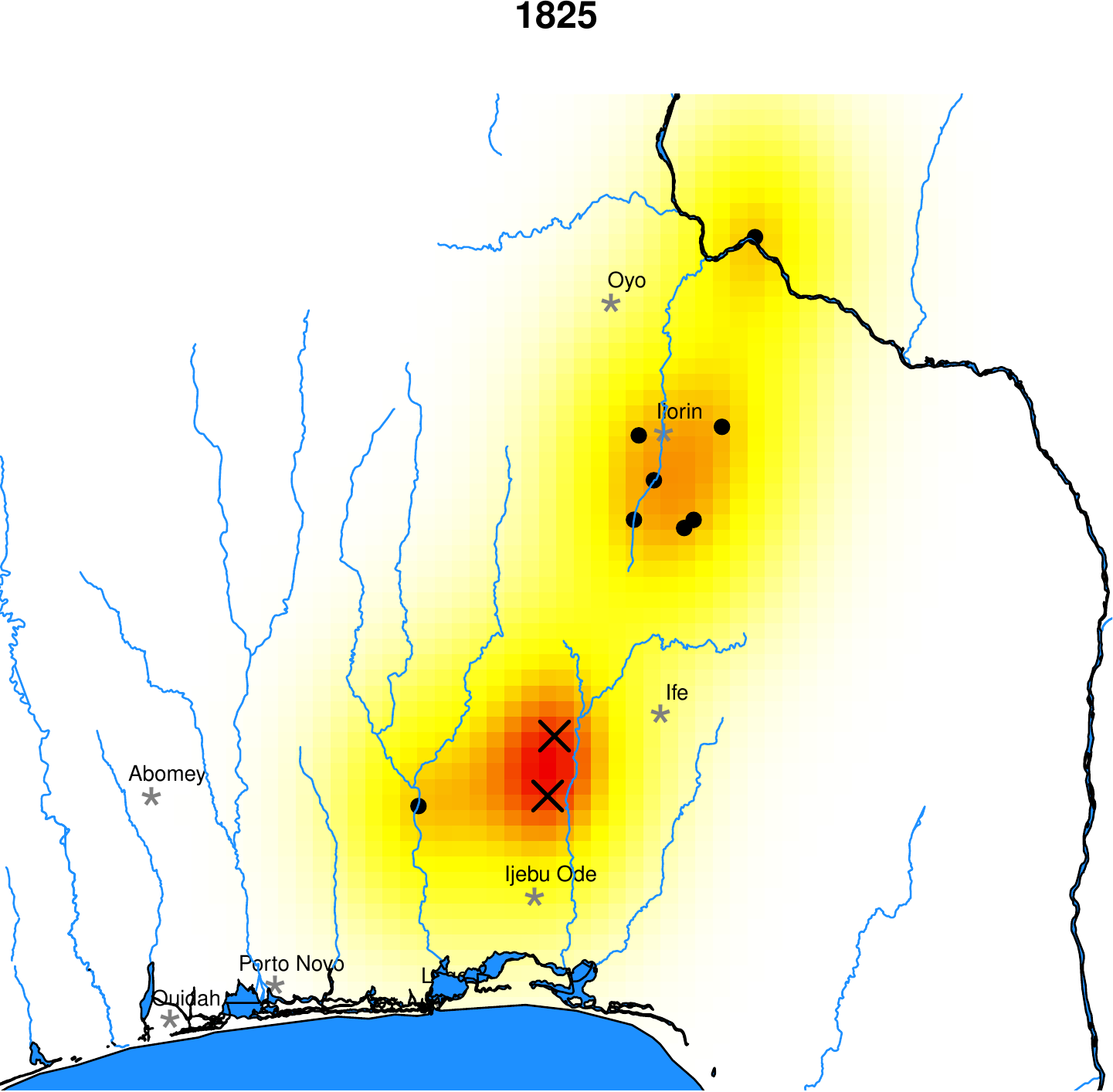}
	\caption{Left to Right: 1825 conflict map via a kernel density estimate with $h=2$; krig with $\nu=.5$ and $a=5/2$; krig with $\nu=3$ and $a=6$}
	\label{fig:krigmodel}
\end{figure}

Figure \ref{fig:krigmodel} shows the flexibility of the Mat\'ern in capturing a shifting border of conflict and contrasts against a kernel density estimate.  The leftmost plot shows a kernel density estimate, where the greater \textit{count} of conflicts in the northeastern area generates a corresponding increased conflict intensity.  The kriging maps on the other hand show the shapes of the conflicts: the parameters $\nu$ and $a$ have some interaction and different combinations can lead to very similar shapes of conflict borders, with differences largely captures in the magnitude of the range parameters.  A larger range - or smaller $a$ - leads to maps with a larger region of uncertainty and non-zero conflict, as shown in the breadth of the yellow region in the middle map compared to the right-most map.

\subsubsection{Estimation}
Rather than provide for fully unconstrained estimation of the 4 spatial covariance parameters, we chose to tune the spatial smoothness according to the heuristics of the underlying problem.  In general, we want to force ranges to be small enough that regions of high correlation are trapped within the same topographical and population areas as the observed conflicts: we found that a $10km$ range accomplished this.  Smoothness was similarly fixed at a 4.5 times differentiable Matern.  While this is fairly smooth in the context of dense spatial data, our observations are quite sparse, and higher smoothness helps ensure the ridge-like structure that mimics a shifting border to reflect the ebb and flow of borders in a region of conventional warfare.  Lower smoothness would enforce a rapid decay to zero away from observations, and would push our model closer to one found from kernel density estimation, as it would result in conflicts being modeled as small, radially symmetric, disconnected, additive kernels about the observed locations.  For the variance parameters of sill and nugget, we fit them via variogram with Cressie weights \citep{cressie1985fitting} to the 1828 data set, then used those parameters to fill in the remaining years. Using these fixed Mat\'ern covariance parameters are all that is required to perform spatial kriging at any desired location.

 In the classical kriging sense, this surface would exist in the units of the $\boldsymbol{Y}$: the 2-valued marker for intensity of battle at a village.  However, we can also view the resulting surface as an implied probability density function, where the higher points of the ridge near conflict locations represent regions of increased probability of slave capture.  By using the kriging estimator to fill in a high-resolution surface over the Oyo region, we then create an annual empirical cumulative density function by dividing the surface by its overall integral.  For each such surface we can simulate sample slave origin locations via direct inversion.

\subsection{Trading Network for Slave Transport}
\label{SS:MDP}
The map in figure \ref{fig:1816TradeMap} displays cities within and around Oyo connected by the most probable trade routes at the time.  Such a depiction naturally translates to a graph-based approach to capturing the economics of the region. From the map, we construct a transition matrix of valid city-to-city movements, and classify a handful of cities where slaves could depart the region: Lagos, Porto Novo, and Ouidah for Atlantic departure; Abomey and Benin City for departure to the neighboring coastal states; and Djougou, Kalama, Bussa, Ogudu, Tsaragi, and Ogodo for departure into the trans-Saharan slave trade.  Rather than directly assign probabilities to the flow of slaves in the region, we turn to a finite horizon Markov decision process to mirror the calculus of slavers at the time.  Markov decision processes (MDPs) can be used for sequential decision making but have been underutilized in the social sciences \citep{boucherie2017markov}.

\subsubsection{A Markov Decision Process}
A Markov decision process describes the partially deterministic and partially stochastic movement of an agent through a network in discrete time. The agent's actions at each state are chosen based on the rewards and costs associated with reaching a state in the network, but the actual event that takes place is probabilistic. Formally, a Markov decision process consists of a 5-tuple $(S, A, P_a, R_a, \gamma)$. $S$ is a finite set of states in a network, often spatially located. $A$ are the actions an agent can take from any given state $s \in S$. $S$ and $A$ can also be thought of as the nodes and edges in a network, respectively. In our case, $S$ corresponds to the cities in the trade network, and $A$ are the valid routes in the trade network a slaver can take; the same set of trade nodes discussed in section \ref{S:Data}. For an action $a \in A$ taken in state $s$, we must define the probability of actually reaching state $s'$ for all states in $S$. Thus, for each action $a$, we must define $P(s_{t+1}' \, | \, s_{t}, a)$. Similarly, we must provide the expected immediate reward/cost incurred after moving from $s$ to $s'$ via action $a$, which we write as $R(s_{t+1}' \, | \, s_{t}, a)$. The idea is that the MDP will designate a best possible route, but a slaver might choose a slightly different one. The difference between the optimal route and the chosen route is reflected in the probabilities $P$, whereas the rewards that determined the best route are saved in $R$. For our purposes, $R$ includes both negative values that represent cost of movement and positive values that correspond to getting a slave to a point-of-sale.  The end result is a "best route" determined by the model for a slaver to reach a point-of-sale, but some chances of deviations along that route to account for the slavers' personal preferences and/or their imperfect information.

The MDP solves the problem of finding an optimal policy $\pi(s)$ whose value specifies the action $a$ to take by the agent at state $s$. The function $\pi$ is found by maximizing some function of the random rewards. Most often this is the expected discounted rewards: 
$$\sum^{\infty}_{t=0} {\gamma^t R_{a_t} (s_t, s_{t+1})}$$
The discount factor $\gamma \in [0, 1)$ allows the rewards incurred in later time steps to be downweighted. The discount factor is fixed at $\gamma=1$ because we found it unnecessary in modeling the historical application. Many algorithms have been developed to solve this optimization problem, e.g. using linear or dynamic programming. We use the policy iteration algorithm implemented in the \texttt{R} package \texttt{MDPtoolbox} \citep{MDPtoolbox}.

Prior to running the MDP, we augment the adjacency matrix $A$ implied by \ref{fig:1816TradeMap} by adding an additional row or state corresponding to each point-of-sale city.  Movement into these added states represents a sale, and holds the postive contents of the rewards $R$.  This flexibility allows some amount of transit between our sale locations - in particular to reflect canoe and naval traffic along the coastal lakes - in order to not view arrival at a port as an inherent end state of the process.  Instead, caravans are free to move until the optimal reward is reached, which may include moving along coasts in the presence of unequal sale rewards.

\subsubsection{MDP Heuristics}
Similar to the choice of kriging, some \textit{ad hoc} decision-making between modeling options is merited.  We harbor multiple criteria for the model for slave transit.  For one, we require sufficient noise or stochasticity to allow for slaves captured in similar locations to deviate in port of departure simply by chance, thought of historically as the personal preferences and knowledge of slavers.  In addition, we require a model with the ability to downweight probabilities of transit based on conflict intensities: in general, slavers would be incentivized to avoid areas of conflict, and as much of the slave transit in the Oyo is state-sanctioned, avoiding regions of conflict also serves as a proxy for slavers tending to avoid the shifting international borders and stay within their preferred home countries.  Finally, it would be ideal for a method to allow for some sale locations to be preferred to others \textit{a priori}, whether that preference is informed by volume of trade or the gradual West-to-East blockade of West African slave ports by the British Navy during the period in question.

This MDP formulation allows for considerable flexibility in meeting our criteria for a transit model. 
\begin{itemize}
\item In an MDP, an optimal choice is calculated, but the underlying Markov chain allows for pseudo-random (non-"optimal") movement from step-to-step.  With a chosen probability, slavers may choose to traverse a sub-optimal route $a$ out of their city/state $s$, possibly including remaining at state $s$ for an additional time-step.
\item Flexibility in rewards can account for both individual slaver preference and the broader temporal shift from the Western Oyo ports and Dahomey to the Eastern ports and Benin.  Setting all point-of-sale rewards to be equal asks the question "what the least resistance route to \textit{any} point-of-sale," whereas varying the reward vector allows for individual slavers to balance preferred or higher revenue sale locations with the implied costs of a longer journey or a journey through regions of conflict.
\item The cost-benefits formulation of MDP reward maximization can also easily be adjusted to downweight specific movements.  In our case, we  explicitly make movement through conflict regions less desirable.  More generally, this formulation could be expanded to include disincentives to cross borders, venture through certain terrains, etc.  Each of these are in addition to the distance-based cost terms we initialize the model with.
\end{itemize}

\subsubsection{MDP Parameter Tuning}
\begin{figure}[t!]
	\centering
	\includegraphics[width=0.45\linewidth]{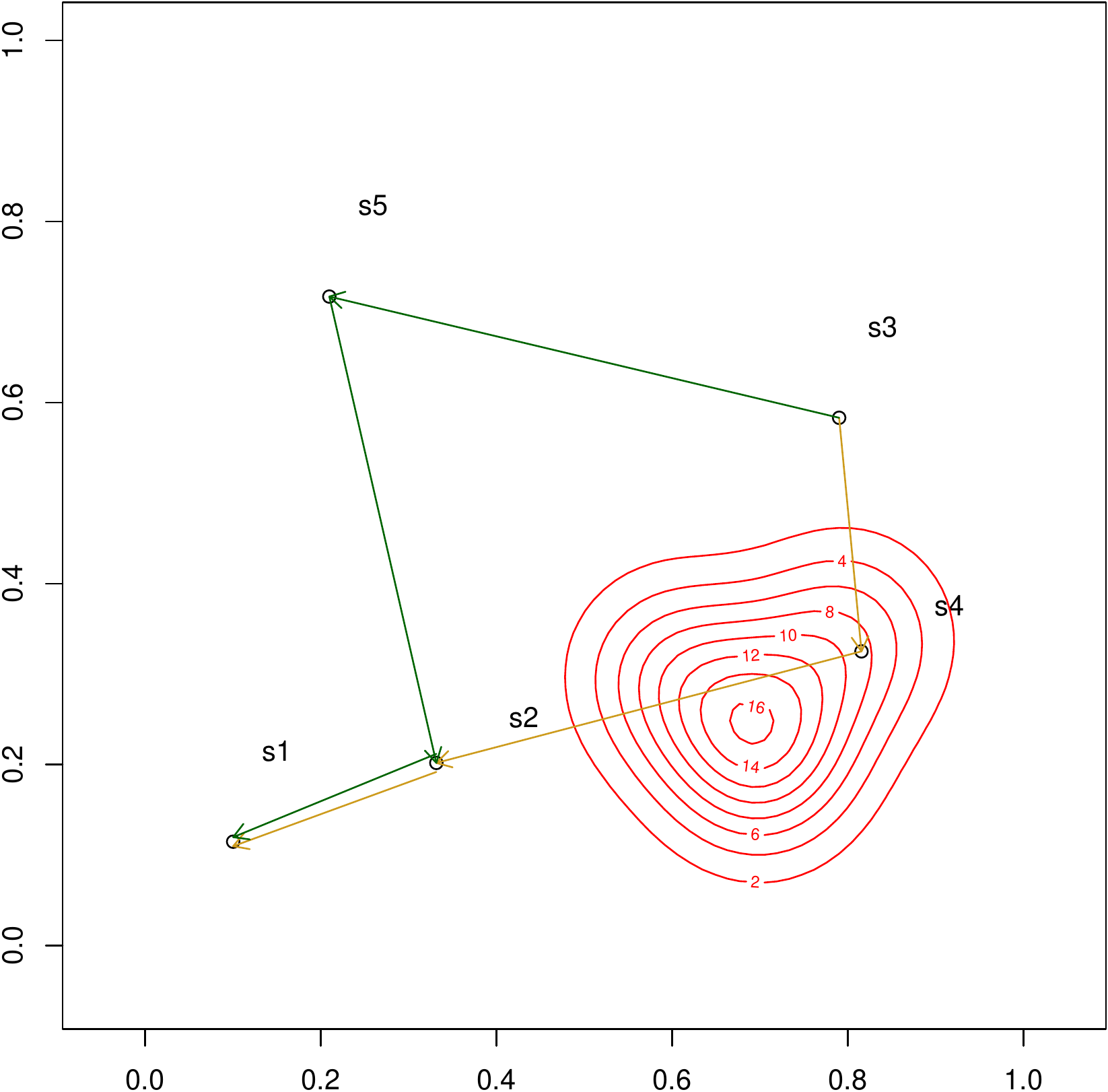}
	\caption{Example of MDP decision chain for a start in S3 with an absorbing state in S1 under no conflict (yellow) or conflict (green)}
	\label{fig:ToyMDP}
\end{figure}

Because we have no information on the price of slaves as a function of point of sale, we choose as a baseline a variant of the Markov decision process with equal expected rewards for each absorbing state.  Transitions along each edge incurred a cost proportional to $D*(1+C)$, where D is the length of that edge and $C$ was the maximum of the conflict kriging predictor along that edge, scaled to an annual maximum of $C=3$.  This allows both the transition chains and the slave origin locations to vary with conflict.  To illustrate this effect, consider the example of a simple MDP seen in figure \ref{fig:ToyMDP}.  In this case, we observe the shortest route being taken in the absence of conflict, but a longer route circumventing the conflict when the intense area of conflict would have overlapped one of the routes taken.

\begin{figure}[t!]
	\centering
	\includegraphics[width=0.45\linewidth]{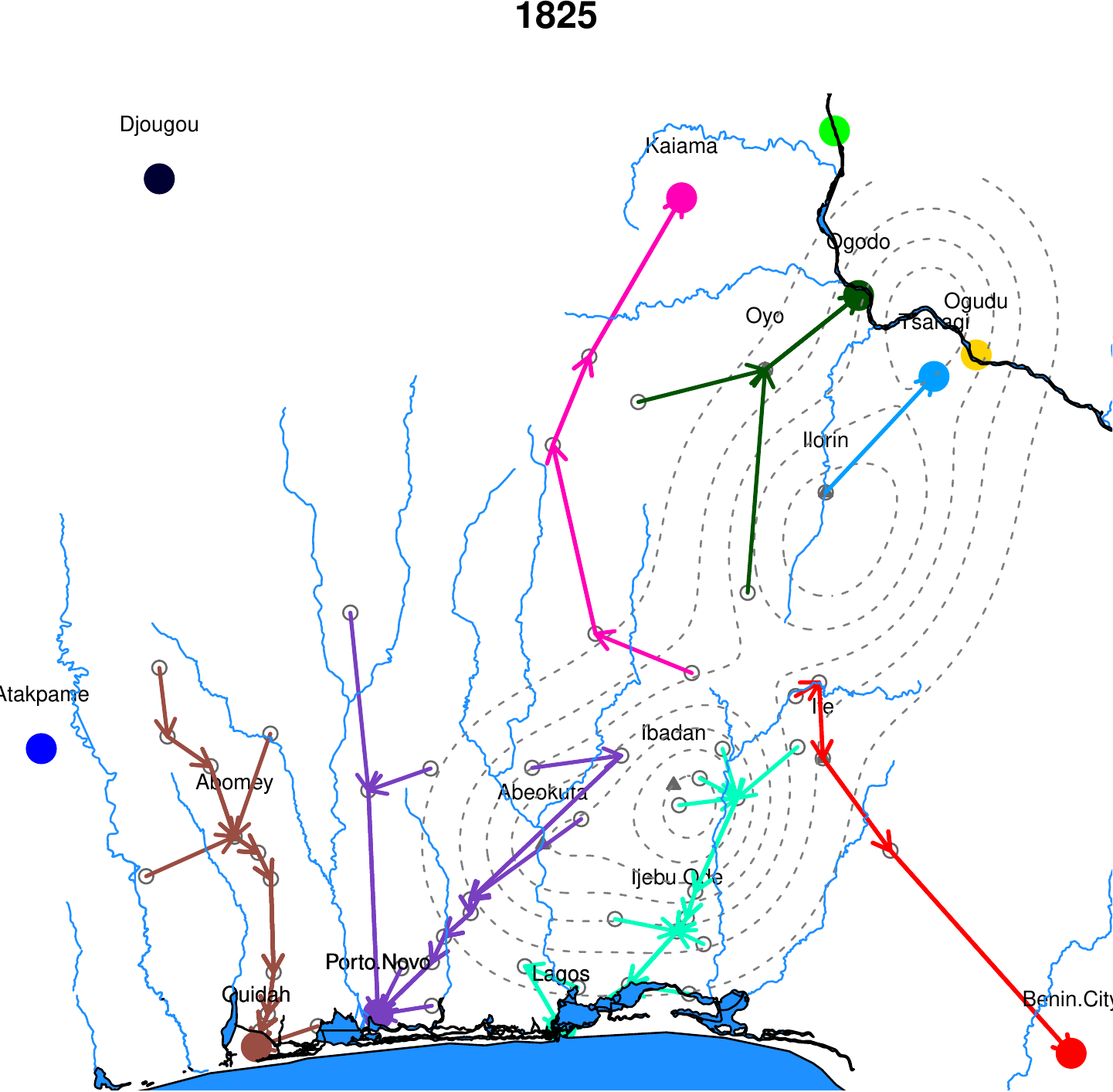}	
	\includegraphics[width=0.45\linewidth]{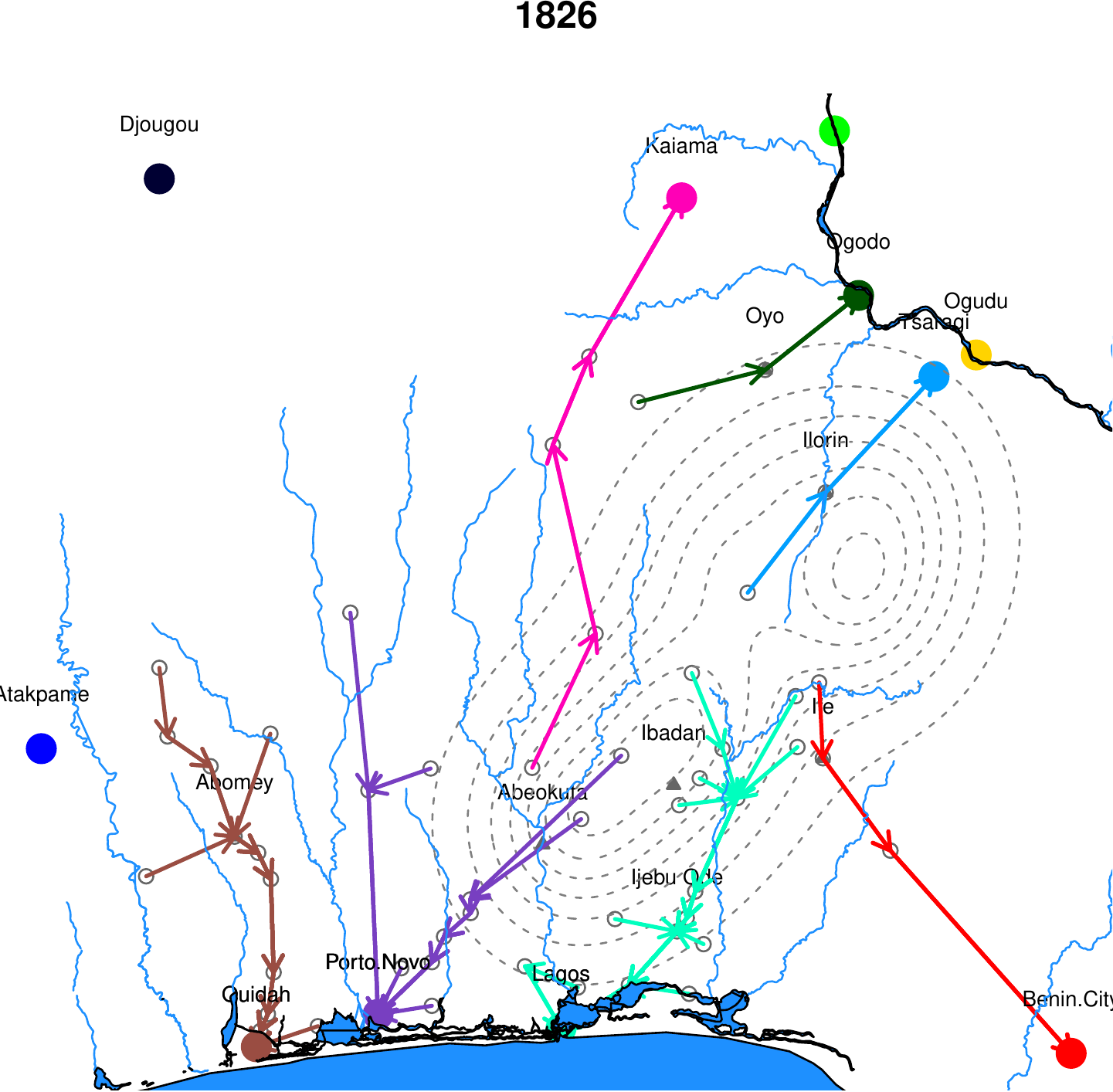}	
	\caption{Example of MDP decision chains for 1825 (left) and 1826 (right)}
	\label{fig:DecisionMaps}
\end{figure}

Figure \ref{fig:DecisionMaps} depicts the decision processes for 1825 and 1826 as sets of arrows connecting each city in the trading network.  In particular, note the difference in decisions made in in the regions around Abeokuta and the label for Ibadan (founded shortly after this conflict).  In general, more trade is flowing northbound in 1826, but we see less traffic through Oyo and Ogodo, instead seeing increases in paths through Ilorin and Kaiama.  Some of this traffic is also deflected away from an eventual port of departure in Porto Novo.

We can combine the kriging conflict estimator with the MDP to simulate the capture of a slave at a specific location and the resulting movement of the slave through the trade network to a point-of-sale.  Use of unequal rewards to vary these simulative results is discussed further in sections \ref{SS:FinalMaps} and \ref{S:Validation}.

\subsection{Maps of Location Given Point-of-Sale}
\label{SS:FinalMaps}
Once the smoothed conflict estimator and MDP process are implemented, repeated simulation of slaves can be passed to the MDP with either varying or identical reward vectors.  This allows for us to create a large sample of slaves and their eventual points-of-sale.  These can be used to describe the ultimate goals: what were the eventual points-of-sale of slaves and also from what conflict and locations did slaves who departed from specific ports originate?

\subsubsection{Large-Scale Simulation}  
\begin{figure}[t!]
	\centering
	\includegraphics[width=0.3\linewidth]{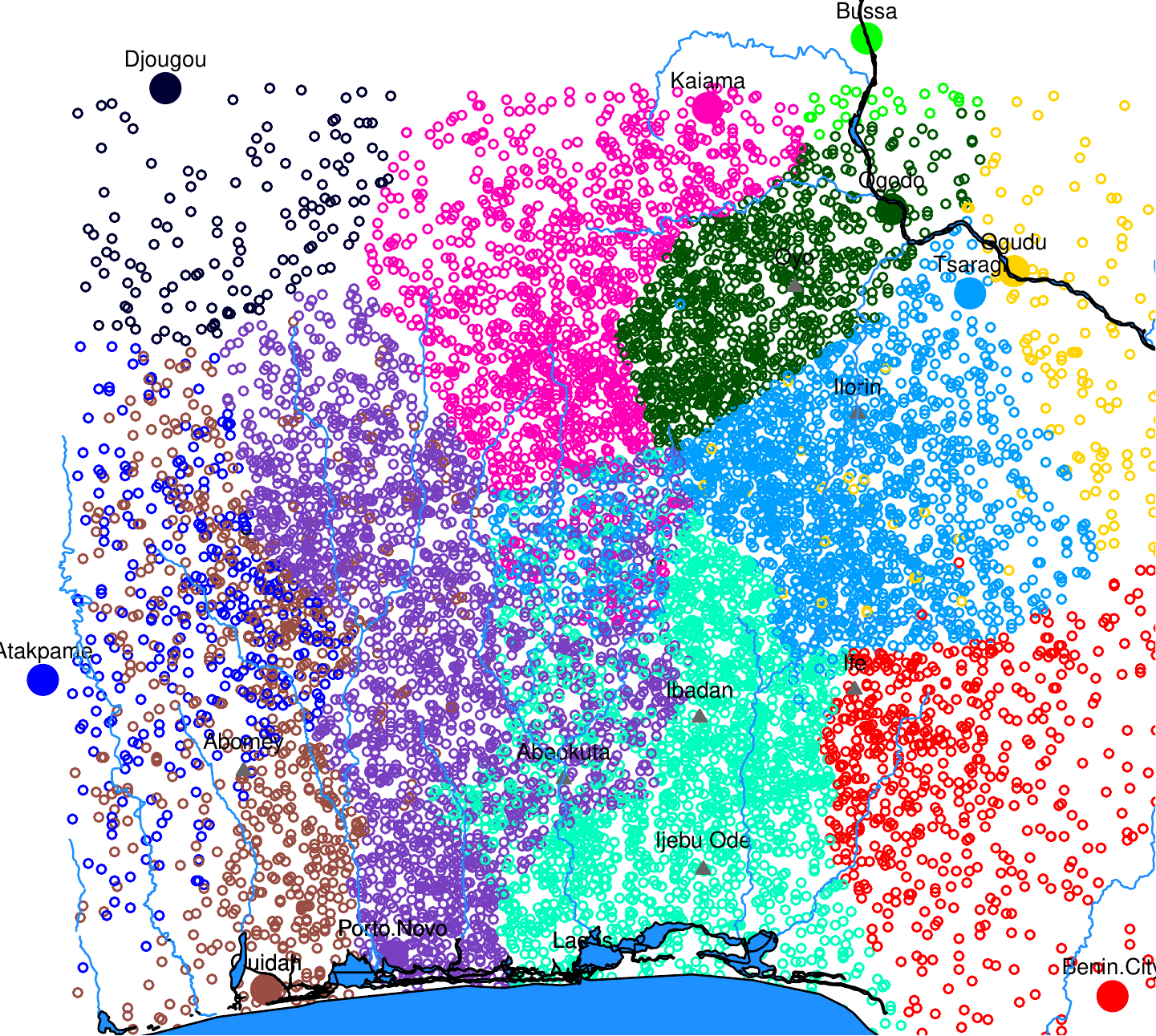}
	\includegraphics[width=0.3\linewidth]{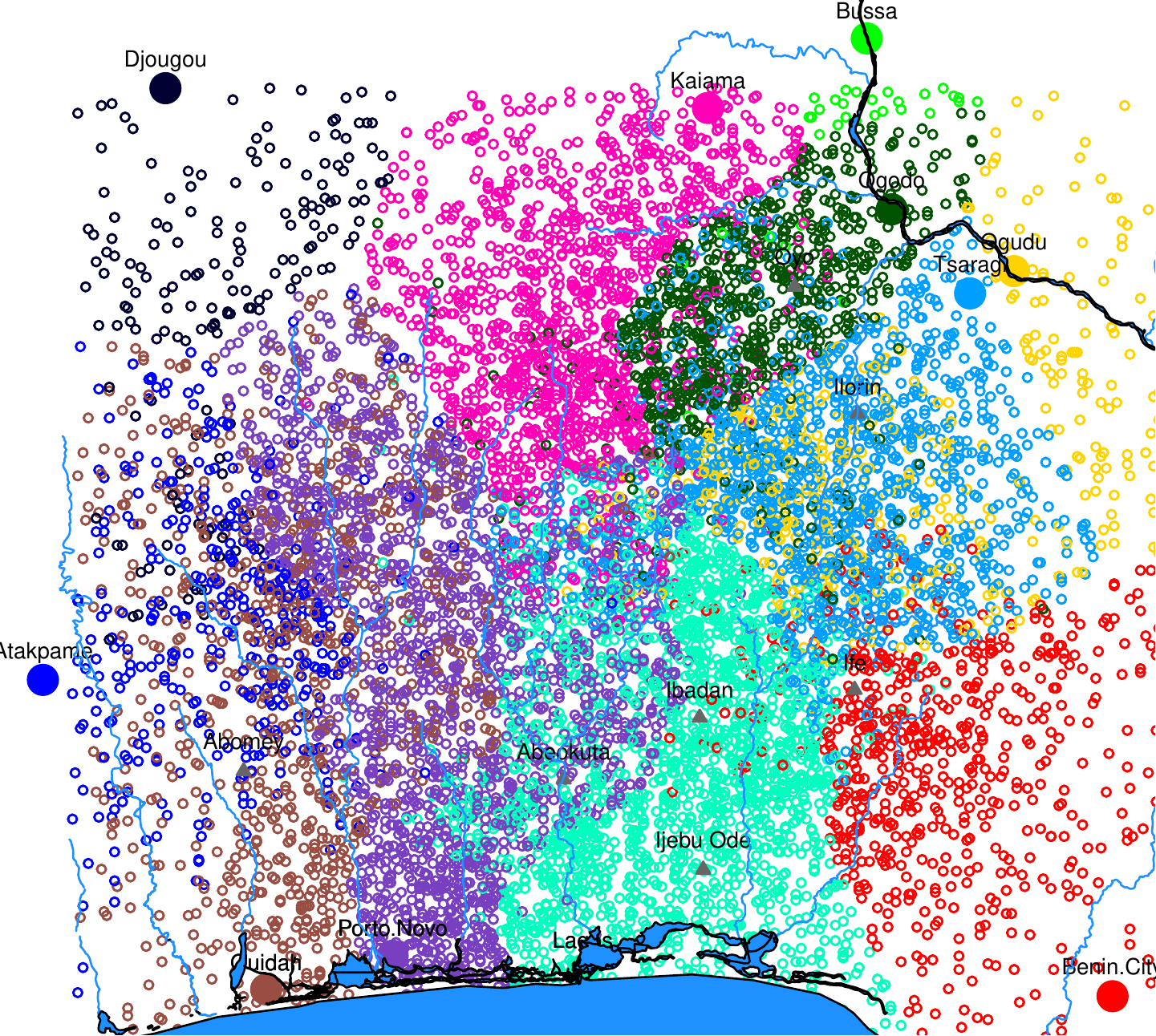}
	\includegraphics[width=0.3\linewidth]{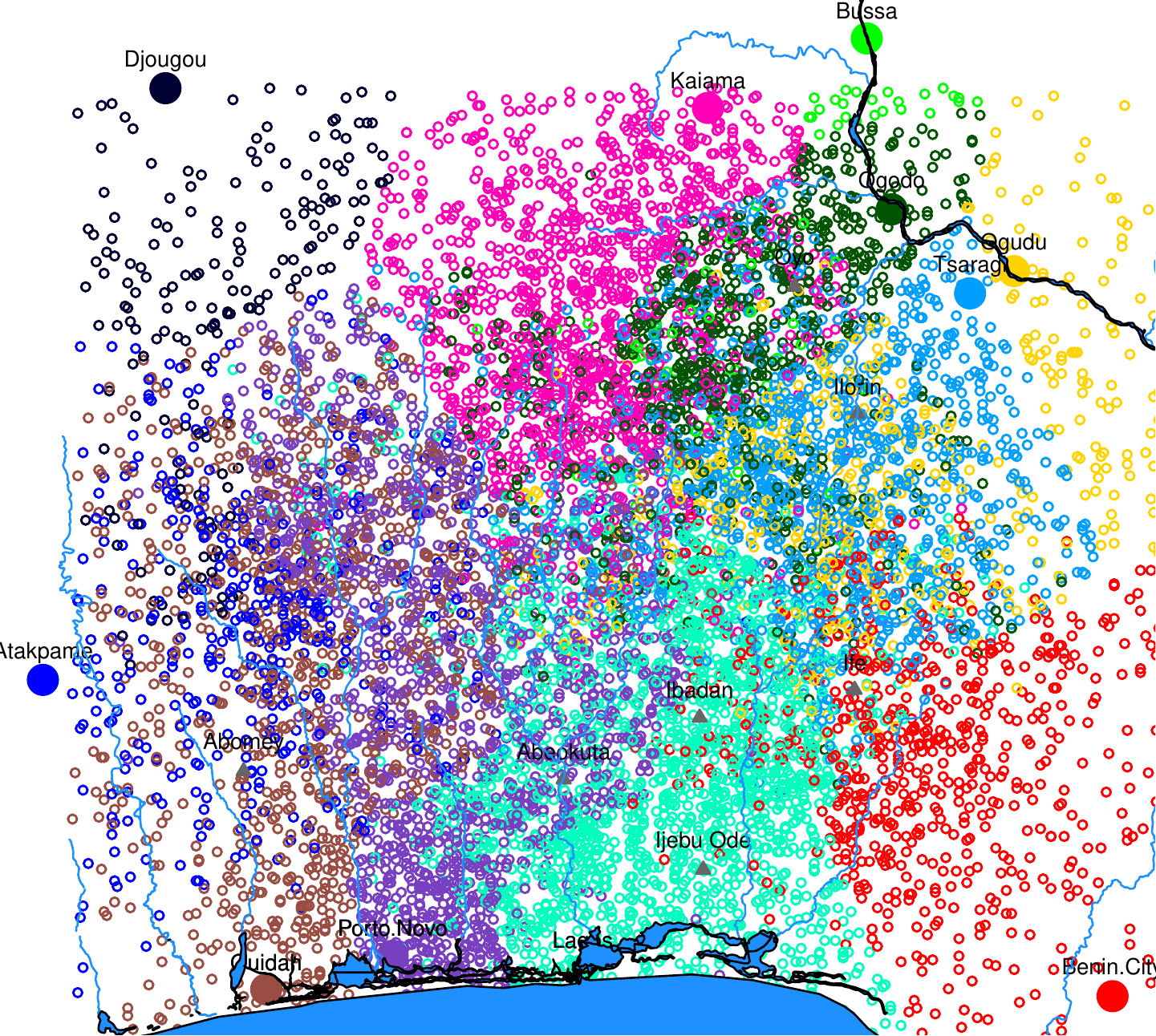}
	\caption{Simulated Slave Origins colored by their points-of-sale.  Left to Right: increasing variance in rewards}
	\label{fig:rewardvar}
\end{figure}

To gain origin and departure information from the models in \ref{SS:krig} and \ref{SS:MDP}, we generate many slaves from direct inversion of the conflict estimator cumulative density functions. Then, for each such slave, we generate a random reward vector from a distribution specifying the end reward for selling a slave at any given absorbing state in the network. Then, we fit an MDP for each individual slave and reward vector pair, resulting in an optimal policy and eventual path of motion for each individual slave. Ultimately, the routing suggested by an MDP with equal rewards at each of point-of-sale is deterministic: every slave caravan of an identical origin location would choose the same optimal route given the annual conflict map.  This doesn't correspond well to the underlying historical narrative: many routes are nearly identical in terms of distance and slavers may have personal connections and preferences leading them to prefer certain routes to others.  Incorporating a random reward vector represents each slaver's knowledge of the conflicts and rewards present, which allows for a non-deterministic result to the question of where a slave will be sold \textit{given} their capture location.  Figure \ref{fig:rewardvar} depicts simulated slaves from the 1832 version of the model with different randomness in the rewards.  With no randomness the boundaries between colors and eventual points-of-sale are strict, whereas the figures shows increasing uncertainty when routes are nearly equivalent in terms of base cost-to-rewards.  Due to the ability to simulate both capture locations and decision processes for any desired number of points, this allows us to generate these mappings to arbitrary probabilistic precision.

\subsubsection{Kernel Density Smoothing For Maps}
One question posed by historians is how to integrate this information - at this point a large collection of origin points encoded by their point-of-departure - into more cleanly interpreted spatial maps. As historians and genealogists often know the port of departure of slaves, we can use the repeated samples of simulated data to figure out such intensity maps for the slaves that left a given port on a given year.  Because the maps will be a set of points corresponds to individual slaves, we are tasked with a similar question as in creating the conflict estimator: how can we provide a continuous image or heat map for slave origin locations given the simulated slaves?  For this we use a simple kernel density estimator, which creates a small, radially decaying kernel function at each simulated slave leaving from a specific port.  The addition of each such kernel function for every slave at a given port gives a heat map for slave origin given port of departure.

Formally, the kernel density estimate takes a radially symmetric function $K(r)$ and estimates the regional heat map $\hat{f}$ via the weighted sum

$$\hat{f}(x)=\frac{1}{nh}\sum_{i=1}^n K\left(\frac{|x-x_i|}{h}\right) $$

where $x_1, x_2, \dots x_n$ are the $n$ slave locations for the slaves departing from the port in question. We choose the multivariate normal as the radial function $K$, as is often convention.  In general, a kernel density estimate requires only tuning one parameter: the bandwidth $h$ that determines the distance/width of the kernel function centered on each simulated slave.  The \texttt{R} function \texttt{kde2d} in package \texttt{MASS} implements the multivariate normal kernel density as its default, and is employed here \citep{MASS}.  While a kernel density function can be sensitive to the number $n$ of points employed, our simulated-based model allows us to simulate any arbitrary amount of spatial samples and construct the resulting kernel density estimator to desired precision.  In the applet mentioned in section \ref{S:App} we allow $h$ to vary from $.5-2$km for a sample of 10,000 simulated slaves and find this provides an appealing map; a larger simulated sample would in turn allow for smaller bandwidths.

\subsubsection{End-Result Parameter Tuning}
\label{SS:Validation}
As laid out so far, our model includes a considerable amount of parameters with no mathematical optimization strategy.  These include the spatial covariance parameters of the kriging estimator, the relative increase in cost of movement to pass through conflict, and the variance in the point-of-sale rewards.

There exist two sources of data that allow for tuning the model to optimize these selections: ship total estimates for ships leaving the southern coast of Oyo and ship ledgers of \textit{names} of slaves as recorded by the colonizing states.

\begin{figure}[t!]
	\centering
	\includegraphics[width=0.6\linewidth]{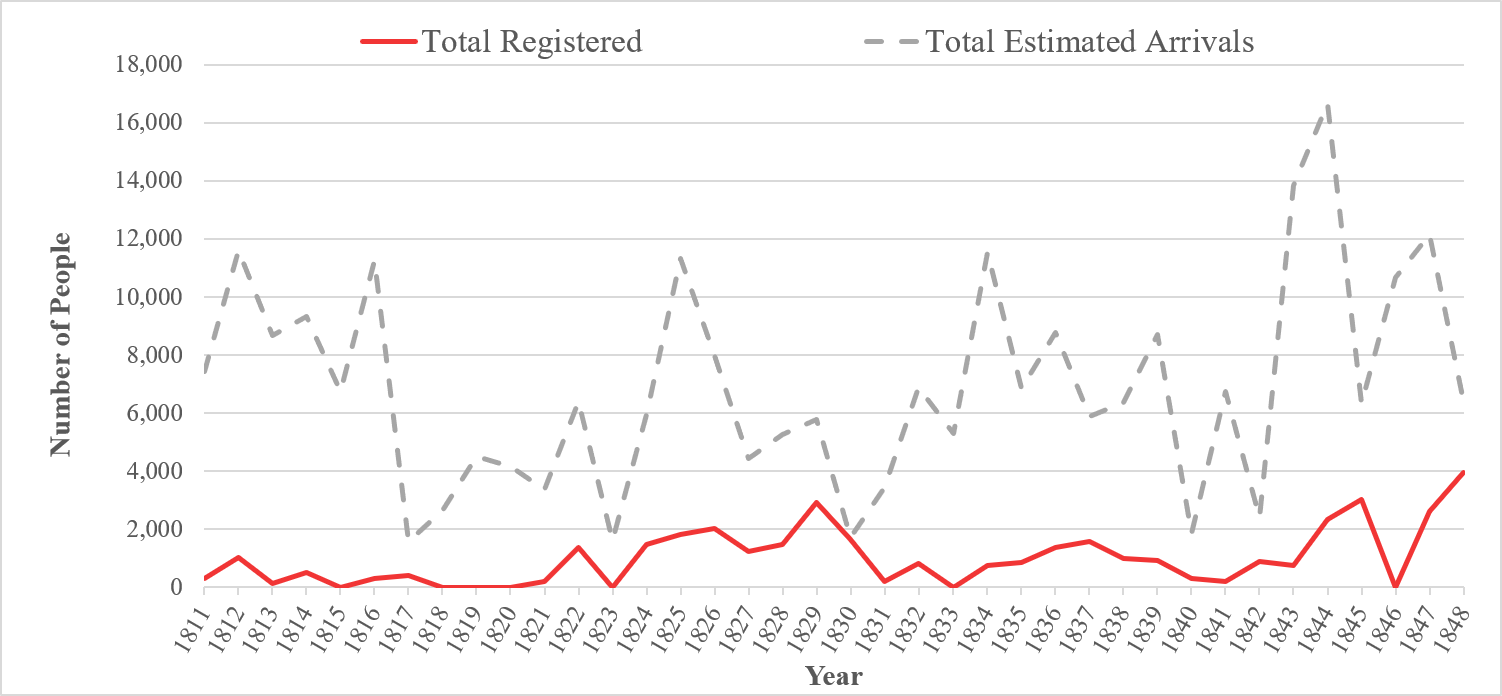}
	\caption{Estimated Versus Recorded Trans-Atlantic Slave Departures from the Bight of Benin}
	\label{fig:MissingShips}
\end{figure}

As presently available to us, the passenger counts on each known ship are considered accurate.  However, as figure \ref{fig:MissingShips} denotes, a considerable amount of slave traffic was not recorded.  Many ships that were recorded in our data set also have no specified ports of departure.  As a result, a considerable amount of geospatial data is missing, and if that data was biased in any way - whether by the British blockades or some other administrative correlation - drawing geospatial conclusions about the within-Africa geospatial data would inherit these biases.

\begin{figure}[t!]
	\centering
	\includegraphics[width=0.45\linewidth]{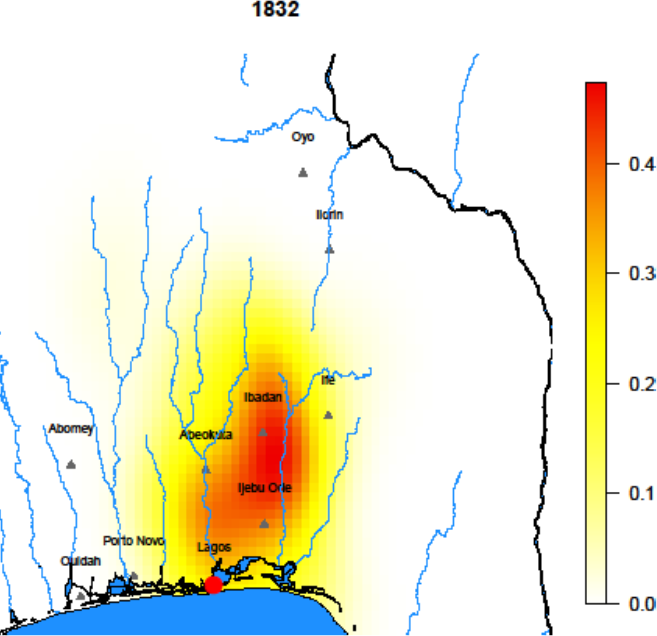}	\includegraphics[width=0.45\linewidth]{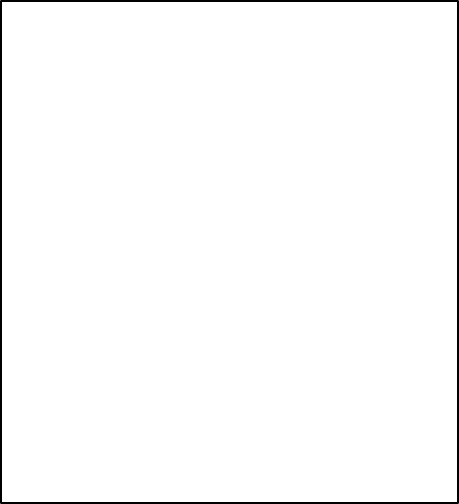}\\
	\includegraphics[width=0.45\linewidth]{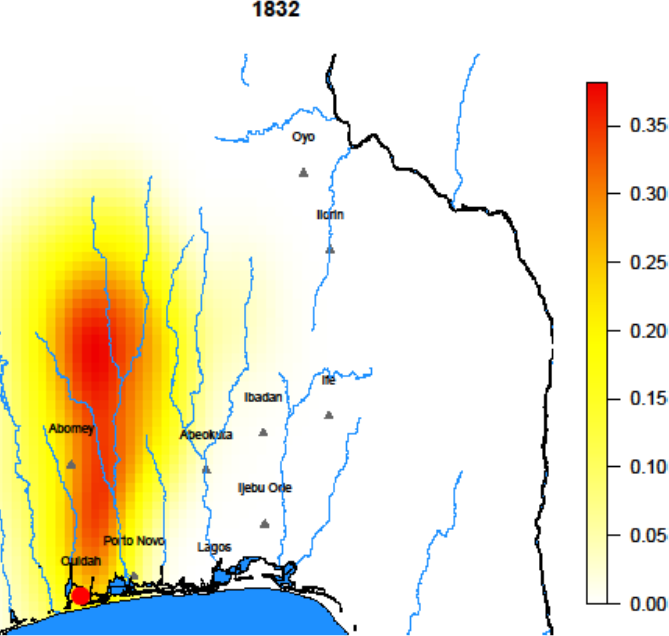}	\includegraphics[width=0.45\linewidth]{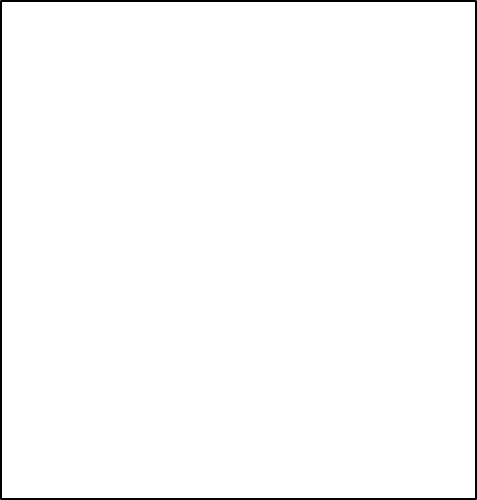}
	\caption{Top: Model and Linguistic Data for a ship leaving Lagos, 1832; Bottom: for Ouidah}
	\label{fig:Linguistic}
\end{figure}

A more recent second stream of data comes from parsing the transcriptions of slave names from known ship logs.  For many ships leaving the area, the Portuguese slave ships recorded each slaves name and attempted to transliterate it into Portuguese.  Recent efforts by historians have begun to translate these names, and placed them into the native tongues of the Bight of Benin.  This allows us to take a few of the ships and have a new kind of mapping: one of general cultural and linguistic origins.  If we take our simulated maps of slave origins, these can be graded and scored against the linguistic data by observing the exact proportions of our simulated data coming from each region.  Such a score specifically suggests a $\chi^2$ optimization, where for each set of parameters we can generate a goodness-of-fit measure that best ensures our simulated departures on each ship most closely match the observed data.  To date, very few ships have been fully transliterated, but a couple such examples coming from 1832 are shown in Figure \ref{fig:Linguistic}.  Here, linguistic distributions are shown as categorical data which must be scored against the plotted distributions.  One advantage of the $\chi^2$ is that each ship added to the linguistic data records can be independently scored as a $\chi^2$, and their additive score is $\chi^2$ as well.

\subsection{Model Summary}
A summary of the model:
\begin{enumerate}
    \item Take space-time locations for conflict and, for each year, create an estimate for the conflict intensity map that represents the shifting border of the wars involved.  This includes some parameters that are viewed as \textit{fixed}.  Simulate slave capture locations out of this heat map.
    \item Create a trade map designating roads, cities, and common caravan routes at the time.  Specify a few cities to be valid points-of-sale in this map.
    \item For each and every slave in step 1, create a Markov Decision Process out of the matrix in step 2 by pairing it with a (randomized) reward vector. Record each slaves point of capture and point of sale.  This includes at least two flexible parameters: the cost-of-movement through conflict $C$ and the variance of the rewards for each sale location.
    \item Aggregate the origin points and points of sale and score the model - either by historians' debate or a $\chi^2$ as linguistic data become available.  Optimize all flexible parameters.
\end{enumerate}

\section{Interactive Web Application}
\label{S:App}
To make this research more widely available to a general audience, we created an interactive web application using the \texttt{Shiny} package in the \texttt{R} programming language. The user can select a year and one or more points-of-sale, and the application generates and displays a conditional probability map showing the most likely region of capture based on our simplified model. The app can also display the yearly conflict data as discrete points, a heatmap of the estimated intensity surface, or a contour plot. Furthermore, the annual approximate state borders \cite{lovejoy2013redrawing} and trade network informing the MDP can be overlayed. 

We have run our model independently for each year from 1816-1836 with the annual trade network and reward vectors changing over time, reflecting the historical narrative. For each year, we generate 10,000 capture locations and record the spatial coordinates, the initial location in the network, and the point-of-sale. We use Kriging on these annual data using the methods in this paper to produce an annual conditional probability surface. For each year, we save the conditional probability Kriging surface, the conflict point data, the KDE conflict intensity surface, the trade network, and the state border shapefiles, which are all the data sets required to host the app.

Our web application is easy to use and freely hosted at \texttt{website.com}. It allows a general audience to interactively explore the history of West African slave trade by visualizing the data and models used in this paper. Note that we do not claim that these maps display the historical truth, but rather the results from a model which provide an approximation of the truth. 

\section{Conclusions and Future Work}
The Markov decision process framework allows for considerable tuning.  While we choose rewards for each slave point of sale that are on average equal between locations, scaling the rewards according to the port departure totals to account for the West-to-East blockading of the region by the British Empire would shift the decision processes of the slavers accordingly.

Our model could also be adapted to account for the time variance in the process and the lack or precision in observed conflict dates. One option would be adding positive spatio-temporal correlation from one year's conflict map to the next.  Another option adds a time delay to each step along the Markov decision process, allowing for recalculations as the conflict shifts each year.

A couple additional sources of validation for potential future work merit discussion.  First, genetic databases and genealogical tracking have become much more powerful in recent years, and we look forward to an increase in the availability of such data.  In particular, if descendents of passengers of any known ships where to compare their genetics to the current genetic mapping of the Bight of Benin areas, we could begin to rapidly improve on the model validation.  A second broader issue our model begins to cover is the difference between traffic through the trans-Saharan and Niger areas.  To date, historians have little understanding or discussion of the amount of slave traffic that moved north out of the coastal regions, and if our model is able to withstand critique for its treatment of the coastal areas, the estimates the MDP-based model gives for northward flow could help provide initial estimates for this movement.

A final source of tuning and validation would be to fit similar models to similar historical situations with better availability of data.  Mapping continuous conflict borders from discrete city observations could be done for nearly any conventional war fought in the $18^{th}$ or $19^{th}$ centuries.  Forced transit situations in the Holocaust did not originate from conventional battles as in Oyo, but have considerably better data due to the relative recency, and could be used to better tune the decision process and exit location models.

We look forward to seeing the expansion of mathematical models in creating both maps in the presence of uncertainty and making those tools available in a non-proprietary form to the public and academic genealogists.
\bibliographystyle{jasa}

\bibliography{Oyobibliography}

	

%

\end{document}